\documentclass[
  prl,
  twocolumn,
  showpacs,
  preprintnumbers,
  amsmath,
  amssymb
]{revtex4}

\usepackage{epsf}
\usepackage{amsfonts}
\usepackage{amssymb}
\usepackage{amsthm}
\usepackage{graphicx}        
\usepackage{dcolumn}         

\def\be{\begin{equation}}
\def\ee{\end{equation}}
\def\bea{\begin{eqnarray}}
\def\eea{\end{eqnarray}}

\newcommand{\ket}[1]{\left|{#1}\right.\rangle}
\newcommand{\mn}[1]{}

\begin{document}

\title{       
Does a Fermi liquid on a half-filled Landau level have Pomeranchuk instabilities?
}
\author{      
  Jorge Quintanilla
}
\email{jorge.quintanilla@stfc.ac.uk}
\affiliation{ 
  ISIS Facility, 
  STFC Rutherford Appleton Laboratory,
  Harwell Science and Innovation Campus,
  Didcot,
  OX11 0QX,
  U.K.
}
\author{      
  Orion Ciftja
}
\email{ogciftja@pvamu.edu}
\affiliation{ 
  Department of Physics,
  Prairie View A\&M University,
  Prairie View,
  Texas 77446, U.S.A.
}

\begin{abstract}
We present a 
theory of spontaneous Fermi surface deformations
for half-filled Landau levels (filling factors of the form $\nu=2n+1/2$). 
We assume the half-filled level to be in a compressible,
Fermi liquid state with a circular Fermi surface.
The Landau level projection is incorporated {\it via} a modified effective electron-electron interaction and the resulting band structure is described within the Hartree-Fock approximation.
We regulate the infrared divergences in the theory and probe the intrinsic tendency of 
the Fermi surface
to deform through Pomeranchuk instabilities. We find that the corresponding susceptibility never diverges, though the system is asymptotically unstable in the $n \to \infty$ limit.
%
%
\end{abstract}
\date{\today}
\maketitle


The discoveries of the integer quantum Hall effect (IQHE)~\cite{klitzing80}
and the fractional quantum Hall effect (FQHE)~\cite{tsui82}
have stimulated many studies on the properties of 
two-dimensional (2D) strongly correlated electronic systems 
in a magnetic field.
New ideas such as 
existence of fractionally charged quasiparticles~\cite{laughlin83} or
the prediction of composite fermions~\cite{jain} 
have broadened our understanding of nature 
and have profoundly affected physics and other sciences.
Theories built on these ideas have also proven very reliable 
to explain unusual properties of strongly correlated electronic systems 
in the extreme quantum Hall regime at various filling factors.

Somewhat more poorly-understood are the half-filled states,
in particular the anisotropic states in
high Landau level-s (LL-s)~\cite{lilly}.
The issue of anisotropy was addressed theoretically quite early on \cite{fogler,1996-Moessner-Chalker}. Moessner and Chalker~\cite{1996-Moessner-Chalker} showed using Hartree-Fock theory that 
a striped charge density wave (CDW) prevails
in the limit of very high LL-s if the interaction is of sufficiently short range relative to the magnetic length. In fact ample theoretical evidence of their existence has accumulated \cite{2001-Shibata-Yoshioka,2003-Goerbig-Lederer-MoraisSmith}.

While a stripe CDW phase is sufficient to explain anisotropy, 
other, competing structures may lead to an anisotropic response as well.
Indeed it was pointed out in Ref.~\cite{1996-Moessner-Chalker} that, 
when the range of the effective interaction is comparable to the magnetic 
length, uniform states may win over the striped phase 
and, a few years later, it was shown \cite{1999-Fradkin-Kivelson} 
that melting of the stripes could lead to a nematic phase. 
In this state, translational  symmetry is completely restored but the system remains anisotropic. 
Interestingly such smectic and nematic states can be regarded
as the `missing links' between the Wigner crystal 
and Fermi liquid states, based on a proposed general picture of 
strong correlations (for a brief 
overview, see \cite{2007-Fradkin-Kivelson-Oganesyan}).

In order to explain the emergence of anisotropy in Fermi
liquid states at filling factor $\nu=2 n+1/2$ $(n \geq 2)$,
there has been a surge of interest in the Pomeranchuk 
instability (PI)~\cite{pomeranchuk}.
Through such mechanism, a compressible Fermi liquid state 
(presumably the half-filled states in high LL-s) may "spontaneosly"
enter into an anisotropic nematic state characterized by a
deformed Fermi surface.
In this scenario anisotropy emerges at one-particle level.
In fact, the wave function for the nematic state proposed in
Ref.~\cite{vadim} consists of single-particle 2D plane-wave states
that form an elliptical Fermi sea~\cite{doan}.
It is also plausible to suggest that anisotropy may 
emerge at a two-particle level, too.
In this second scenario, one would start with a 
broken rotational symmetry wave function
that contains a suitable symmetry-breaking parameter in the
two-particle correlation part of the wave function
and a Slater determinant of 2D plane waves that form a 
standard 
circular Fermi sea~\cite{2002-Ciftja-Wexler,ijmpb}. 


In this Letter we test the first of the above two scenarios. Specifically, we address the question of formation of a nematic 
state (or its higher angular momentum generalisations) 
from the opposite direction in the phase diagram 
of Ref.~\cite{2007-Fradkin-Kivelson-Oganesyan}, 
i.e. from the Fermi liquid side. 
Assuming a circular Fermi liquid ground state as a starting point,
our {\it modus operandi} is to derive an effective electron-electron 
interaction that takes into account the projection onto the 
relevant half-filled LL. 
We then investigate whether this circular Fermi surface is unstable to small, 
point group symmetry-breaking deformations that are the 2D counterpart
of the three-dimensional (3D) PI~\cite{pomeranchuk}.
%
%

As was discussed more generally in 3D, for several model effective 
interactions~\cite{2006-Quintanilla-Schofield}, 
key requirements to have a PI are: 
(i) a sharp feature in the interaction potential at 
    some characteristic distance; and 
(ii) for this characteristic length to be larger than the separation distance 
     between particles in the system. 
Indeed the range of the effective 
interaction potential that we derive is longer than the magnetic length. 
Moreover it has non-monotonic features that develop into a sharp kink 
at a particular distance that increases as the order of the LL-s increases. 
On the other hand, the average distance between particles in a half-filled LL 
is fixed by the mangetic length. 
Thus both of the above conditions are met asymptotically 
for sufficiently high LL-s. Testing the ensuing expectation of an intrinsic tendency to a PI for high half-filled LL is the main purpose of this work.


We consider filling factors of the form:
$\nu=2 \, n+ \nu^{\ast}$ where $n=1, 2, \ldots$ 
is the index of the uppermost half-filled LL
($\nu^{\ast}=1/2$) assumed to be fully spin-polarized. 
According to the Halperin-Lee-Read theory \cite{HLR}, 
the $N^{\ast}$ electrons of the half-filled LL form a 2D 
circular Fermi liquid state
(with density $\rho^{\ast}$) 
that effectively sees no magnetic field. 
We have $\nu^{\ast}= 2\pi l_0^2 \rho^{\ast}$
where the magnetic length $l_0 \equiv \sqrt{\hbar c/eB}$
represents the characteristic length scale in the problem.
Since $\nu^{\ast}=1/2$ and 
$\rho^{\ast}=\pi \, k_F^2/(2 \, \pi)^2$,
the radius of the circular 2D Fermi wave vector is $k_F=1/l_0$.
%
%
%
Assuming that kinetic energy is quenched 
the Hamiltonian, up to a constant, is
\be
\label{eq:vfield}
\hat{H}_{n}=\frac{1}{2} \int d^2r \int d^2r^{\, \prime} 
\rho_{n}(\vec{r}) \, v(|\vec{r}-\vec{r}^{\, \prime}|) \,
\rho_{n}(\vec{r}^{\, \prime}) \ ,
\ee
where $v(|\vec{r}-\vec{r}^{\, \prime}|)=
e^2/(4 \, \pi \, \epsilon \, |\vec{r}-\vec{r}^{\, \prime}| )$
is the Coulomb potential.
For simplicity we will take the dielectric function $\epsilon$ to be a constant. 
Here
$\rho_{n}(\vec{r})=\Psi_{n}^{\dagger}(\vec{r}) \Psi_{n}(\vec{r})$
is the density operator. The quantum field operator, 
$\Psi_{n}^{\dagger}(\vec{r}),$
creates a spinless electron in the $n^{\mbox th}$ LL at position $\vec{r}$.

It is well known that projection onto the $n^{\rm th}$ LL
contained implicitely in the definition of 
the density operator $\rho_n(\vec{r})$ 
introduces highly non-trivial physics in the Hamiltonian in 
Eq.~(\ref{eq:vfield}).
The key idea of our approach is to assume a uniform circular Fermi liquid state and investigate possible instabilities
induced 
by the heavily renormalised interactions.

One can transform the density operator in 2D Fourier space as 
$\rho_n(\vec{q})=\int d^2r \, e^{i \, \vec{q} \, \vec{r}} \, \rho_n(\vec{r})$
and rewrite Eq.(\ref{eq:vfield}) as
$\hat{H}_n=\frac{1}{2} \sum_{\vec{q}} v(q) \,
 \rho_n(-\vec{q}) \, \rho_n(\vec{q})$
where
$v(q)={2 \pi e^2}/(4\pi\epsilon q)$
is the 2D Fourier transform of the Coulomb interaction potential.
Based on the Hamiltonian theory of Ref.~\cite{murthyshankar}
we can project the density onto the half-filled LL, which amounts to writing
$\rho_n(\vec{q})=F_n(q) \, \overline{\rho}(\vec{q})$
where 
$\overline{\rho}(\vec{q})$ 
is the projected density operator.
Thus the Hamiltonian can be rewritten in 2D Fourier space as
\be
\hat{H}_n=\frac{1}{2} \sum_{\vec{q}} V_{n}(q) \,
  \overline{\rho}(-\vec{q}) \, \overline{\rho}(\vec{q}) \ ,
\label{hamiltonian}
\ee
where
$V_{n}(q)=F_n(q)^2 \, v(q)$
represents an effective interaction potential
which in real space would be given by
\(
  V_n(r) = \int d^2q/(2 \, \pi)^2 \, e^{-i \, \vec{q} \,\vec{r}}\, V_n(q)
\).
%
%
%
The form factors are
$F_n(q)=L_n\left( q^2 l_0^2/2 \right) \exp \left(-q^2 l_0^2/4 \right)$ 
where the $L_n(x)$-s are Laguerre polynomials.  
%


There are two consequences of the projection onto the half-filled Landau level implicit in the Hamiltonian (\ref{hamiltonian}): firstly, the effective interaction potential $V_n(r)$ is heavily renormalized when compared to the Coulomb interaction. Secondly, the projected density operators themselves have a non-trivial algebra. Here we wish to investigate whether the first of these two features may be sufficient to produce a PI instability in a putative Fermi liquid state. In line with this view, we assume that new fermion creation and annihilation operators $\psi^{\dagger},\psi$ can be introduced in such a way that the projected densities can be written as $\bar{\rho}(\vec{q})=\sum_{\vec{k}}\psi(\vec{k})^{\dagger}\psi(\vec{k}+\vec{q})$ (a form that implies standard operator algebra). Here $\psi^{\dagger}(\vec{k})$ creates a fermion in a plane wave state. One can then apply the mean-field theory of a PI in a 3D continuum in Ref.~\cite{2006-Quintanilla-Schofield} (recently generalised to 2D \cite{2008-Quintanilla-Haque-Schofield,2007-Quintanilla-Hooley-Powell-Schofield-Haque}). It starts with a Hartree-Fock ansatz for the ground state, 
\(
   \ket{\varepsilon_{\vec{k}}} = 
   \prod_{\vec{k}} 
   \left[
      \Theta\left(
      	 \varepsilon_{\vec{k}}
      \right)
      +\Theta\left(
      	  -\varepsilon_{\vec{k}}
      \right)
      \psi^\dag(\vec{k}) 
   \right]
   \ket{0},
   \label{2}
\)
where $\ket{0}$ is the vacuum. Note that this ansatz is a homogeneous, itinerant state with an arbitrary dispersion 
relation $\varepsilon_{\vec{k}}$. 
For the Hamiltonian in (\ref{hamiltonian}), 
this Slater determinant of plane waves affords a rudimentary description 
of the re-emergence of itinerancy when the kinetic energy 
has been completely quenched (note that, unlike Refs.~\cite{2006-Quintanilla-Schofield,2008-Quintanilla-Haque-Schofield,2007-Quintanilla-Hooley-Powell-Schofield-Haque}, 
there is no `bare' contribution to $\varepsilon_{\vec{k}}$ here - 
it all comes from interactions). 
The functional form of $\varepsilon_{\vec{k}}$ is our variational parameter. 
It determines which plane wave states are 
occupied {\it via} $\varepsilon_{\vec{k}} \leq 0$ and is found by minimization 
of   
$\langle \hat{H}_n \rangle$. 
That yields a self-consistency equation giving the shape of the Fermi surface 
through $\varepsilon_{\vec{k}}=0$. 
PI-s are point group symmetry-breaking instabilities 
of the shape of the Fermi surface.

To find an instability equation, it is natural to split 
$\varepsilon_{\vec{k}}$ in two parts: 
one that preserves the continuous rotational symmetry of the plane and 
another one that may break it. 
We thus write \cite{2008-Quintanilla-Haque-Schofield}
\be
\varepsilon_{\vec{k}}
=
\varepsilon_{0}(|{\vec{k}}|)-\Lambda_l(|\vec{k}|)\cos(l \, \theta_{\vec{k}}),
\label{X}
\ee
where 
\(
  \varepsilon_{0}(|{\vec{k}}|) 
\)
is the symmetric component of the dispersion relation and $l=1,2,3,\ldots$ 
determines the symmetry of the instability. 
The condition of instability towards a small deformation of the Fermi surface 
is (ignoring the possibility of a first-order phase transition) 
\cite{2007-Quintanilla-Hooley-Powell-Schofield-Haque,2008-Quintanilla-Haque-Schofield}
\begin{eqnarray}
  V_l
 & \geq &
  \frac{4\pi\hbar v_F^0}{k_F^0},
  \label{inst_eq_gen} 
  \\
  \mbox{ where } V_l & =  &
  4\pi\int_0^{\infty}dr ~ r \, V_n(r) \,  J_l(k_F^0 \, r)^2 
  \label{Vl_rspace}
\end{eqnarray}
measures the strength of the electron-electron interaction in the channel 
with angular momentum $l=1,2,3,\ldots$ 
For $l=1$, the PI corresponds to a rigid displacement 
of the Fermi suface in reciprocal space, 
without change of either shape or volume. 
This can never lead to a lowering of the energy in a Galilean-invariant 
system and therefore this instability cannot take place \cite{2006-Quintanilla-Schofield,2006-Wolfle-Rosch,2008-Quintanilla-Haque-Schofield}.
[This also follows from the more explicit form of the instability equation 
derived below, Eq.(\ref{inst_eq_vF0}).] 
On the other hand for $l=2,3,4\ldots$ we could have PI-s corresponding 
to deformations of the Fermi 
surface possessing $d$-wave, $f$-wave, $g$-wave... symmetry, respectively. 
The Fermi velocity in Eq.~(\ref{inst_eq_gen}) 
is given by \cite{2008-Quintanilla-Haque-Schofield}:

\be
  v_F^0 = 
  \frac{k_F^0}{4\pi\hbar} V_1,
  \label{vF0_sc}
\ee
where we take into account the
infinite bare mass $m \to \infty$ of 
our Hamiltonian [see Eq.~(\ref{hamiltonian})].
Substituting this into the instability equation (\ref{inst_eq_gen}) yields
\be
  V_l-V_1
  \geq
  0.
  \label{inst_eq_vF0}
\ee

\begin{figure}
\begin{center}
  \includegraphics[width=0.9\columnwidth]{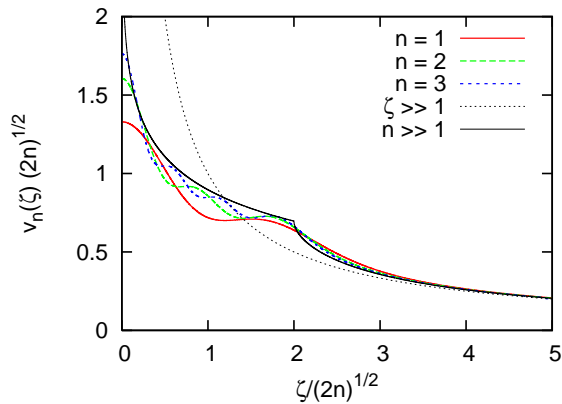}
\end{center}
\caption{\label{F1}
        Interaction potentials 
        calculated from Eq.(\ref{eq:vnreal2})
        alongside the large-$n$ [Eq.~(\ref{vn_large_n})] and 
        large-$\zeta$ (Coulomb) asymptotes.}
\end{figure}

The effective interaction can be written as
\be
\label{eq:vnreal2}
V_n(r)=\frac{e^2}{4\pi\epsilon} \int_{0}^{\infty} dq \, J_0(q \, r)
\left[ L_n\left( \frac{q^2 l_0^2}{2} \right) \right]^2 
       \exp \left(-\frac{q^2 l_0^2}{2} \right) \ ,
\ee
where $J_l(x)$ are Bessel functions and $n=1,2,3,\ldots$ 
%
%
%
We can write
$V_n(r)=(e^2/4\pi\epsilon l_0) \, v_n(\zeta)$
where $\zeta=r/l_0$
is the natural dimensionless distance.
%
%
We calculated $v_n(\zeta)$ exactly for several $n$
(we do not display such expressions for the sake of brevity).
The results are shown in Fig.~\ref{F1}, where
we also plot an asymptotic expression 
for the effective interaction obtained in the limit of high LL-s: \cite{2003-Goerbig-Lederer-MoraisSmith}
\be
  v_n\left(\zeta\right)
  \sim
  \frac{1}{\zeta}
  \frac{4}{\pi^2}
  \Re
  \left[
    K
    \left(
      \frac{1}{2}-\sqrt{\frac{1}{4}-\frac{2n}{\zeta^2}}
    \right)^2
  \right]  \ \  ; \ \
  n \gg 1.
  \label{vn_large_n}
\ee
$K(x)$ is the complete elliptic integral fo the first kind.
%
%
%
%
%
%
%
%
%
%
All interaction potentials tend symptotically to the Coulomb interaction,
\(
  v_n(\zeta) \sim {1}/{\zeta}
  \label{vn_large_zeta}
\)
for large $\zeta$.
This leads to infrared divergences in the theory whose regulation is 
discussed below. 
On the other hand, the presence of a length scale in the problem 
(the magnetic length) is apparent at shorter distances where
one notices that as $n$ increases an increasingly sharp kink 
develops at the specific distance $r_n = \zeta_n l_0$ where
\be
  \zeta_n \equiv 2\sqrt{2n} \ .
  \label{znd}
\ee 
Note in particular that the ultraviolet divergence of the 
Coulomb interaction at short distance has been suppressed. 
Interestingly, a sharp feature in the interaction potential at a 
finite distance $r_n$ suggests the 
possibility of a PI provided the 
dimensionless parameter $r_n k_F$ is large 
enough \cite{2006-Quintanilla-Schofield}. 
Since $r_n k_F = \zeta_n$, 
Eq.~(\ref{znd}) implies an increased tendency towards a PI in high LL-s.

\begin{figure}
\includegraphics[width=1\columnwidth]{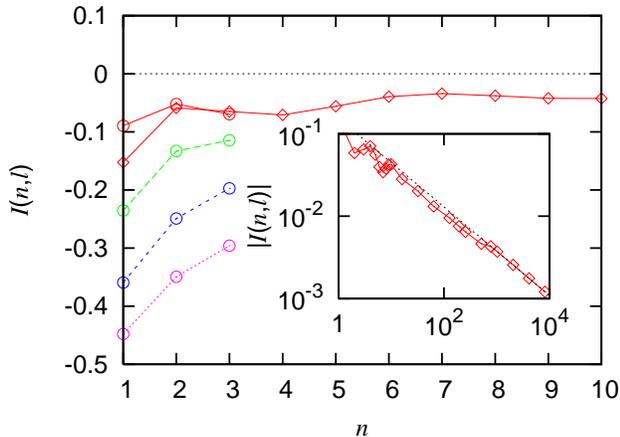}
\caption{\label{fig:conv3}
         Dependence of the integral $I(n,l)$ in Eq.~(\ref{in_eq_exp_2}), 
         evaluated numerically, on $n$. 
         The angular momenta of the instabilities are $l=2,3,4\mbox{ and }5$
         (solid, long-dashed, short-dashed and dotted lines, respectively). 
         Circles represent values obtained using the effective
         potentials, $v_n(\zeta)$ for $n=1, 2$ and $3$.
         Diamonds were obtained using the asymptotic 
         formula in Eq.(\ref{vn_large_n}). 
         The inset shows the behaviour at very large $n$ for 
         the $l=2$ case. 
         The straight line in the inset is a fit of the form
         $I(n,l) = A_l/n^{B_l}$
         to the numerical data for $n \geq 800$.
         The fitting parameters are 
         $A_2 \approx 0.15$ and $B_2 \approx 0.54$.}
\end{figure}

Note that since $v_{n}\left(\zeta\right) \sim 1/\zeta$ for $\zeta \gg 1$ 
the RHS of Eq.~(\ref{Vl_rspace}) diverges. 
This leads to diverging Landau parameters, 
$F_l \propto V_{l}$ \cite{2008-Quintanilla-Haque-Schofield} and 
Fermi velocity $v_{F}^{0} \propto V_1$. 
These infrared divergences can be regulated by introducing 
a cutoff $\zeta_c$ in the electron-electron interaction potential. 
The actual values of $v_F$ and $F_l$ depend on the actual value 
and form of the cutoff which in turn depends 
on \emph{extrinsic} features of the system such as 
sample size or device configuration. 
However, as long as the cutoff distance is long enough, the 
instability condition does not depend on these parameters. 
It is such \emph{intrinsic} tendency to a PI 
that we wish to probe further. 
%
To do so let us rewrite the instability condition in Eq.(\ref{inst_eq_vF0})
in a more explicit form: 
\begin{multline}
I(n,l) \equiv \int_0^{\infty}d\zeta \, \zeta \, v_n(\zeta) \,
  \left[ J_l(\zeta)^2-J_1(\zeta)^2 \right] \geq 0 \ .  
  \label{in_eq_exp_2}
\end{multline}
Clearly, if $I(n,l) < 0$, a PI does not occur. 
However, as $I(n,l)$ becomes less negative, the susceptibility to 
a PI with angular momentum quantum number $l$ 
increases until it diverges at $I(n,l)=0$.
%
%
%
%
%
\begin{figure}
\includegraphics[width=1.0\columnwidth]{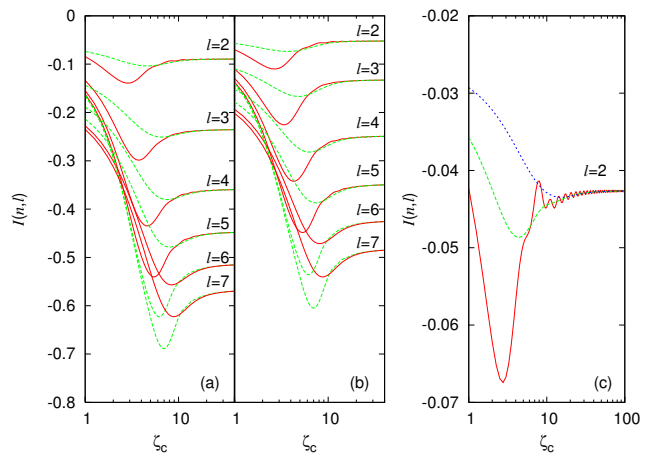}
  \caption{\label{fig:conv2}\label{fig:conv2_asymp}
            Dependence of $I(n,l)$ in Eq.~(\ref{in_eq_exp_2}) on the
            cutoff $\zeta_{c}$ for 
            $w=1$ (red solid curves), $2$ (green long dash) and 
            $4$ (blue short dash) 
            for (a) $n=1$; (b) $n=2$; and (c) $n=10$ 
            (note the different scales on the third plot). 
            We employed the asymptotic expression in Eq.~(\ref{vn_large_n})
            to plot (c).
            The values of angular momentum $l$ are as indicated.}
\end{figure}
The following replacement
\begin{equation}
v_{n}\left(\zeta\right)\to f\left(\frac{\zeta-\zeta_{c}}{w}\right)
v_{n}\left(\zeta\right),
\label{cut}
\end{equation}
with $f\left(x\right)=\frac{1}{e^{x}+1}$
provides a mathematically convenient way of introducing a cutoff in the 
form of a smoothed-out step function of finite width $w$. 
Fig.~\ref{fig:conv2} shows the dependence of $I(n,l)$ on the 
cutoff $\zeta_c$. 
For sufficiently large $\zeta_c$ the integral is independent of $\zeta_c$ 
as well as of the width $w$ of the cutoff. 
Taking the limit $\zeta_c \to \infty$ represents a convenient way
to estimate the \emph{intrinsic} tendency of the system to a PI 
which is independent of the form and value of the cutoff. 
Physically, such a cutoff may correspond to, for example, 
the finite thickness of the device.
The crucial point is that
the value of the integral $I(n,l)$ in Eq.~(\ref{in_eq_exp_2}) 
is an \emph{intrinsic} feature of the system 
independent of the form, size and mechanism of the cutoff
as long as $\zeta_{c}$ is much larger than the average
separation between particles $\sim k_{F}^{-1}=l_{0},$ 
and $\zeta_{c}\gg \zeta_n$ where $\zeta_n$ is the
distance where the kink 
of the effective potential 
in Fig.~\ref{F1} appears. 
%
%
The dependence of the converged values of $I(n,l)$ on $n$ and $l$
is shown in Fig.~\ref{fig:conv3}. 
%
%
We note that the PI condition is never met for $\nu=5/2,\,9/2$ and $13/2$.
The susceptibility to a PI has a non-trivial dependence on $n$ with a 
maximum for $l=2$ at $n=2$, while it monotonically increases 
for $l>2$ up to $n=3$. 
Also note that the system seems most susceptible to a PI
in the $l=2$ channel. 
%
%
From these results we conclude that the system never has an intrinsic 
PI (i.e. $I(n,2)$ is always $<0$). 
On the other hand, for large $n$ we have $I(n,2) \propto 1/n^{0.54}$, 
which implies that the half-filled level is asymptotically 
unstable to a PI in the $n \to \infty$ limit \footnote{The strict $n \to \infty$ limit corresponds to zero magnetic field, where focusing on an isolated, half-filled LL is not justified. This result must therefore be interpreted as an enhanced tendency towards a Pomeranchuk instability as $n$ is increased.}.

\emph{In summary}, 
we have investigated whether a compressible circular Fermi liquid state
in half-filled high LL-s undergoes a phase transition to a non-circular
anisotropic nematic phase through a PI.
We use the Hamiltonian theory approach~\cite{murthyshankar} to derive a properly
LL projected effective electron-electron interaction from which itinerancy 
emerges at the mean-field level. 
We look for deformations of the Fermi surface by testing for 
an \emph{intrinsic} divergence of the corresponding susceptibility.
We find that the susceptibility towards a PI is increasingly large 
as we move to high LL-s and diverges in the $n \to \infty$ limit. 
The increased tendency towards a PI is a 
direct consequence of the length scale $r_n \sim \sqrt{2 \, n} \, l_0$ 
present in the effective interaction. 

As is well known, there is also a tendency towards stripe 
formation \cite{fogler,1996-Moessner-Chalker} 
with whom the PI competes. 
In real systems, there is a second length scale, $r_c$, 
responsible for cutting off infrared diveregences associated 
with the long-range nature of the Coulomb interaction. 
Evidently $r_n$ and $r_c$ would become comparable for large values of $n$.
The intrinsic susceptibility to a PI is already quite high in this regime. 
Thus, even minor \emph{extrinsic} effects 
(sample thickness, device configuration, etc.) associated
with this second length scale may trigger such an instability.
A calculation of the critical value of $n$ at which this transition
might take place is beyond the scope of the present analysis. It would require taking device configuration into account as well as comparing the energy of any Fermi liquid state to those obtained for stripe configurations.

We acknowledge useful discussions with R. Rold{\'a}n and N.I. Gidopoulos. J.Q. acknowledges financial support from CCLRC (now STFC) in association with St. Catherine's College, Oxford. O.C. acknowledges financial support from the National Science Foundation Grant No. DMR-0804568. 


\end{document}